\documentclass[12pt]{article}
\def\ad{{\rm ad\,}}
\def\bra{\langle}

\def\diag{{\rm diag\,}}

\def\ket{\rangle}
\def\offdiag{{\rm offdiag\,}}
\def\dist{{\rm dist\,}}
\def\dom{{\rm dom\,}}
\def\gtrless{\raise2.5pt\hbox{$>$}\!\!\!\!\lower2.5pt\hbox{$<$}}
\def\id{{\rm id\,}}

\def\N{{I\!\!N}}
\def\nid {\noindent}
\def\QED{\mbox{\rule[-1.5pt]{6pt}{10pt}}}
\def\R{{I\!\!R}}
\def\Ran{{\rm Ran\,}}
\def\spect{{\rm spect\,}}
\def\tvert{{|\!|\!|}}
\def\vsth{\vskip 2mm}
\def\Z{Z\!\!Z}

\def\cA{{\cal A}}
\def\BB{{\cal B}}

\def\EE{{\cal E}}
\def\HH{{\cal H}}
\def\KK{{\cal K}}
\def\OO{{\cal O}}

\begin{document}
\begin{center}
{
{\bf\huge
Progressive diagonalization and applications}$^\spadesuit$}
\vskip5mm
P. Duclos$^{\sharp,\flat}$, O. Lev$^\natural$,
\v S\v tov\'\i\v cek$^\natural$, M. Vittot$^\sharp$
\vskip5mm

$\sharp$ Centre de Physique Th\'eorique, Marseille

$\flat$ PhyMat, Universit\'e de Toulon et du Var

$\natural$ Depart. Math. Fac. of Nucl. Sc., Czech Technical University,
Prague

$\spadesuit$ talk given at the $4^{\rm th}$ Operator Algebras Conference, "Operator
Algebras \& Mathematical Physics", Constan\c ta, July 2001

\end{center}

Abstract. We give a partial review of what is known so far on stability of
perio\-dically driven quantum systems versus regularity of the bounded driven
force. In particular we emphasize the fact that unbounded degeneracies of the
unperturbed Hamiltonian are allowed. Then we give a detailed description of an
extension to some unbounded driven forces. This is done by representing the
Schr\"odinger equation in the instantaneous basis of the time dependent
Hamiltonian with a method that we call progressive diagonalization.

\section{The main theorem}

This paper concerns the spectral analysis of Floquet Hamiltonians
associated to quantum systems which are periodically driven. They are described
by the Schr\"odinger equation:
\begin{equation}\label{schrodingerEquation}
\left(-i\partial_t+H_0+V(\omega t)\right)\psi=0,\qquad
\left\{\matrix{H_0\ {\rm selfadjoint\ in}\ \HH,\hfill\cr
 t\to V(t),\ 2\pi\ {\rm periodic,}\cr
 \omega>0,\ \mbox{a real frequency,}\hfill\cr
 \R\ni t\to \psi(t)\in\HH,\hfill\cr}\right.
\end{equation}
where $\HH$ is a separable Hilbert space, and
$H_0$ has the following type of spectral decomposition
($E_n$, $P_n$ denoting respectively the eigenvalues in ascending order and the
eigenprojections)
$$
H_0=\sum_{n=1}^\infty E_{n}P_n,\qquad M_{n}:=\dim P_n<\infty
$$
with a growing gap condition of the type
\begin{equation}\label{growingGapCondition}
\exists\sigma>0,\qquad{1\over \left(\Delta E_\sigma\right)^\sigma}:=
\sum_{m\ne n}
{M_{m}M_{n}\over|E_{m}-E_{n}|^\sigma}<\infty.
\end{equation}

The driven force is given by a time dependent real potential $V$ which is,
in the first part of this paper, bounded in the following norm
\begin{equation}\label{normOfV}
\|V\|_r:=\sup_{m\in\N}\sum_{k\in\Z}\sum_{n\in\N}\|V(k,m,n)\|\max\{|k|^r,1\}.
\end{equation}
where $\|V(k,m,n)\|$ denotes the operator norm of
\begin{equation}\label{componentOfV}
V(k,m,n):={1\over2\pi}\int_0^{2\pi}e^{-i k t}P_{m}V(t)P_{n}dt:\HH\to\HH.
\end{equation}
The following main theorem is about the selfadjoint operator $K:=K_0+V$ with
$K_0:=-i\omega\partial_t\otimes 1+1\otimes H_0$ acting in the Hilbert space
$\KK:=L^2(S^1)\otimes\HH$, i.e. functions which are $2\pi$-periodic in time.
\vsth

\nid{\bf Theorem~1}.

{\sl\ Let $\omega_0>0$, $\Omega_0:=[{8\over
9}\omega_0,{9\over8}\omega_0]$, assume (\ref{growingGapCondition}) for some $\sigma>0$
and let
$$
\Delta_0:=\min_{m\ne n}|E_m-E_n|.
$$
Then,
$
\forall r>\sigma+{1\over2},\quad \exists C_1>0\ {\rm and}\ C_2(\sigma,r)>0,
$
such that
$$
\|V\|_r<\min\left\{{4\Delta_0\over C_1},\, {\omega_0\over C_1},\, {\omega_0\over
C_2}\left(\Delta E_\sigma\over\omega_0\right)^\sigma\right\}\quad \Longrightarrow
$$
$$
\exists\Omega_\infty\subset \Omega_0,\quad{\rm with}\quad
{|\Omega_\infty|\over|\Omega_0|}\ge 1-{\|V\|_r\over {\omega_0\over C_2}{\left(\Delta
E_\sigma\over\omega_0\right)^\sigma}}
$$
so that $K$ is pure point for all $\omega\in\Omega_\infty$. $|\Omega_\star|$
denotes the Lebesgue measure of $\Omega_\star$.}
\vsth
\nid The proof of this theorem and its complement that we state at the end
of this section can be found in \cite{DLSV}. This theorem is a result in singular
perturbation theory since as this is shown in \cite{DSV} one has
$$
(\ref{growingGapCondition})\quad\buildrel{\rm
obviously}\over\Longrightarrow\quad\limsup_{n\to\infty}E_n=+\infty
\quad\buildrel\mbox{\cite{DSV}}\over\Longrightarrow\quad\forall
a.a.\
\omega,\ \spect K_0=\R
$$
i.e. for almost all $\omega$, $K_0$ has a dense pure point spectrum. To be able
to overcome this small divisors difficulty we use a technique which consists in
applying to $K_0+V$ an infinite sequence of unitary transforms so that at the
$s^{\rm th}$ step
$$
K_0+V\sim K_0+G_s+V_s,\quad{\rm with}\quad
V_s=\OO(\|V\|_{r-\sigma-{1\over2}}^{2^{s-1}})
$$
i.e. $K_0+V$ is unitarily equivalent to a diagonal part $K_0+G_s$, in the eigenbasis
of $K_0$, plus an off diagonal part $V_s$ which is super exponentially small in
the
$s$ variable provided $\|V\|_r$ is small enough. This is why we like to call this
method {\sl progressive diagonalization} although it is known usually under the
name KAM-type method, since this is an adaptation of the famous
Kolmogorov-Arnold-Moser method originally invented to treat perturbations of
integrable Hamiltonians in classical mechanics.

An extension of the previous theorem to certain classes of unbounded perturbations
$V$ is given in section \S3, see Theorem~3. We shall do it by ( block-)
diagonalizing
$H_0+V(t)$ for each
$t$, i.e. by constructing a time dependent unitary transform $J(t)$ such that
$H_0+V(t)=J(t)(H_0+G(t))J(t)^\star$, where $H_0+G(t)$ commutes with $H_0$, thus
$$
K_0+V\sim -i\omega\partial_t +H_0 +G(t)-i\omega J(t)^\star \dot J(t)
$$
($\dot J$ denotes the time derivative of $J$). $V$ and $H_0$ are such that the new
perturbation
$G(t)-i\omega J(t)^\star
\dot J(t)$ is bounded so that we can apply Theorem~1. This
diagonalization of
$H_0+V$ will be done in details with a progressive diagonalization method (P.D.M.),
however simpler than the one used for Theorem~1 since we do not have small
divisors here. We think it's a good starting point for readers which are not
familiar with this P.D.M.. This idea of regularizing an unbounded $V$ by going
to the instantaneous basis of $H_0+V(t)$ is not new, ( see e.g. \cite{H3,ADE}).
Let us mention the recent work \cite{BaG} which also treats the Schr\"odinger
equation with unbounded perturbations which are quasi-periodic and
{\em analytic} in time; here we treat the {\em differential} periodic case.

The use of KAM technique to diagonalize quantum Floquet
Hamiltonians appeared first in \cite{B} who considered pulsed rotors of the type
\begin{equation}\label{pulsedRotor}
-i\omega\partial_t+H_0+f(t)W(x)\quad{\rm acting\ in}\quad L^2(S^1)\otimes
L^2(S^d),
\end{equation}
where $d=1$, $H_0=-\Delta$, $f$ and $W$ are analytic. Later on, the adaptation of
the Nash-Moser ideas to treat non analytic perturbations was done in \cite{C} for the
special case of (one dimensional) driven harmonic oscillators. These ideas
where extended to a large class of models in
\cite{DS}. However to our knowledge the above Theorem~1 is the first one who
allows degeneracies of eigenvalues of $H_0$ which are not uniformly bounded
with respect to the quantum number $n$. Consequently we can exhibit frequencies
such that the quantum top model in arbitrary dimension, i.e. the higher
dimensional versions of the pulsed rotor (see (\ref{pulsedRotor}) and
\S4.1), is pure point.

One of the main goal of the spectral analysis of these Floquet Hamiltonians is the
study of the stability of periodically driven quantum systems since it is known that
\begin{equation}\label{RAGEStability}
\mbox{$K_0+V$ is pure point}\buildrel\mbox{\cite{EV}}\over\iff
\lim_{n\to\infty}\sup_{t\gtrless0}\|\sum_{m=n}^\infty P_m\psi(t)\|=0,\quad \forall
\psi(0)\in\HH
\end{equation}
because $\exp(-iT(K_0+V))$ is unitarily equivalent to $1\otimes U(T,0)$ where $U(T,0)$
denotes the propagator over the period $T$ associated to the Schr\"odinger equation
(\ref{schrodingerEquation}), (see \cite{H1,Y}). The above r.h.s. says that the
probability that the quantum trajectory with an arbitrary initial condition $\psi(0)$
explores in the full history the eigenstates of $H_0$ of energy higher than
$E_n$ becomes smaller and smaller as $n$ gets larger and larger. On the other
hand if
$\psi(0)$ belongs to the continuous spectral subspace of $U(T,0)$ then (see \cite{EV})
$$
\forall m\in\N,\qquad\lim_{t\to\infty}{1\over t}\int_0^t\|P_m\psi(t)\|dt=0
$$
which means that in the time average the probability that the trajectory stays
in the
$m^{\rm th}$ spectral subspace of $H_0$ vanishes.

The conclusion that can be drawn from the articles \cite{B,DS,DLSV} is that for non
resonant ( i.e. diophantine) frequencies the pulsed rotor is stable if the driving
force is sufficiently regular in time ( see Figure~1 below) and sufficiently small in
amplitude. In addition it is known, (see \cite{EV}) that if
$f$ is sufficiently regular in time and $\omega$ is resonant, i.e. rational, the
pulsed rotor is stable. The situation is
different for the kicked rotor ( i.e.
$
f(t):=\delta(t)
$, the Dirac distribution): it has been proven in \cite{CaG} that if the
frequency is rational or even Liouville one can find $W$'s such that $U(T,0)$
has a continuous spectral component. However nothing is known for non resonant
frequencies. Since the kicked rotor corresponds to $r<-1$ in the notation of
(\ref{normOfV}) and the known values of $r$ for which $U(T,0)$ is pure point
are
$r>3/2$ the sequences of papers
\cite{B,DS,DLSV} can be considered as reports of the efforts devoted to the long
march from the pulsed rotor to the kicked rotor ( in the non resonant case). In
Figure~1 below we give a diagram which tells the history of this march.
Since the regularity in the space variable has also played a role we present
this diagram in the plane of points $(r_1,r_2)$ which say that the following
generalization of (\ref{normOfV})
$$
\|V\|_{r_1,r_2}:=\sup_{m\in\N}\sum_{k\in\Z}\sum_{n\in\N}
\|V(k,m,n)\|\bra k\ket^{r_1}\bra m-n\ket^{r_2}
$$
is finite, with $\bra x\ket^2:=1+x^2$.

\vskip15mm

\begin{center}
\setlength{\unitlength}{0.25mm}
\begin{picture}(300,300)
\put(0,0){\vector(1,0){300}}
\put(300,-10){\mbox{$r_1$}}
\put(0,0){\vector(0,1){300}}
\put(-15,300){\mbox{$r_2$}}
\put(220,0){\line(0,1){5}}
\put(225,10){\mbox{\small analyticity}}

\put(0,220){\line(1,0){5}}
\put(-60,260){\mbox{\small analyticity}}
\thicklines
\put(220,220){\line(1,0){80}}
\put(220,220){\line(0,1){80}}
\put(225,240){\small\shortstack[c]{KAM\\
 Pulsed Rotor\\
 \cite['85]{B}}}
\thicklines
\put(100,300){\line(0,-1){200}}
\put(100,100){\line(1,-2){50}}
\thinlines
\put(150,0){\line(0,1){5}}
\put(149,10){\mbox{\small 17}}
\put(100,0){\line(0,1){5}}
\put(93,10){\mbox{\small 10}}
\put(150,150){\shortstack[c]
 {Nash-Moser\\
 \cite['96]{DS}}}
\put(100,100){\line(1,0){200}}
\put(150,30)
{\small\shortstack{Nash-Moser\\
+ Adiabatic Regularization\\
 \cite['96]{DS}}}
\put(0,100){\line(1,0){5}}
\put(-20,95){\mbox{10}}
\thicklines
\put(15,0){\line(0,1){300}}
\put(10,-15){\mbox{$3\over2$}}
\put(18,80){\small\shortstack{
 Nash-\\Moser\\
 + tricks\\
\cite['01]{DLSV}}}
\thinlines

\put(-18,228){\large\mbox{$\times$}}
\put(-20,200){\vector(1,3){10}}
\put(-90,190){\small\mbox{kicked rotor}}
\put(-25,-15){\mbox{\small$-1$}}
\put(-10,0){\line(0,1){5}}
\put(-20,0){\line(1,0){20}}
\put(-18,18){\vector(1,-1){18}}
\put(-50,20){\mbox{\small$(0,1)$}}
\put(-80,320){\small\mbox{$\sigma=1+0$,
\ $\sup_n{M_n}<\infty$, \
$\omega$ non
resonant and
$\|V\|_{r_1,r_2}$ small enough}}
\put(25,10){\mbox{$5\over2$}}
\put(25,0){\line(0,-1){25}}
\put(25,-25){\line(1,0){270}}
\put(150,-15){\small\mbox{this paper}}
\put(0,0){\line(0,-1){35}}
\put(-23,-30){\small\mbox{$-1$}}
\put(0,-25){\line(1,0){5}}
\put(0,-25){\line(1,0){1}}
\put(3,-25){\line(1,0){1}}
\put(6,-25){\line(1,0){1}}
\put(9,-25){\line(1,0){1}}
\put(12,-25){\line(1,0){1}}
\put(15,-25){\line(1,0){1}}
\put(18,-25){\line(1,0){1}}
\put(21,-25){\line(1,0){1}}
\put(24,-25){\line(1,0){1}}
\end{picture}

\vsth\vsth\vsth\vsth
Figure~1. Historical diagram of progress toward the kicked rotor
\end{center}
\vsth
The pure pointness of $K$ from which follows the stability (\ref{RAGEStability})
does not imply in general that
\begin{equation}\label{dynamicalStability}
\sup_{t\gtrless0}(H_0\psi(t),\psi(t))<\infty
\end{equation}
i.e. the uniform boundedness of the energy. Notice that the converse is obviously
true. It is believed that to get (\ref{dynamicalStability}) one should require
sufficient regularity of the eigenprojectors of $K$. That is why the following
complement to theorem~1 may be of interest. We have also added some explicit
bound on the constants $C_1$ and $C_2$. It will be necessary in \S3 to consider
potentials
$V$ which depend on the frequency $\omega$ in a more elaborate way.
Suppose that
$V:\R\times\R_+\to\BB(\HH)$ is a bounded measurable function, which is $2\pi$ periodic with
respect to the first variable and such that for almost all $t\in\R$ and $\omega\in\R_+$,
$V(t,\omega)^\star=V(t,\omega)$. For such $V$ me modify $\|V\|_r$ as follows:
$$
\|V\|_r:=\sup_{\omega,\omega'\in\Omega_0}\sup_{m\in\N}\sum_{k\in\Z,n\in\N}
\left(\|V_{kmn}(\omega)\|+\omega_0\|\partial_\omega
V_{kmn}(\omega,\omega')\|\right)\max\{|k|^r,1\}
$$
where
$$
V_{kmn}(\omega):={1\over2\pi}\int_0^{2\pi}e^{-ikt}P_mV(t,\omega)P_ndt
$$
and
$$
\partial_\omega
V_{kmn}(\omega,\omega'):={V_{kmn}(\omega)-V_{kmn}(\omega')\over\omega-\omega'}.
$$
\vsth

\nid{\bf Complement to Theorem~1}.

{\sl\ In addition to the statements of Theorem~1 one also has
\vskip1mm
\nid (a) each
eigenprojection $P$ of $K$ is bounded in the norm
$$
\|P\|_{r-\sigma-{1\over2}}=\sup_{m\in\N}\sum_{k\in\Z}\sum_{n\in\N}
\|P(k,m,n)\|\max\{|k|^{r-\sigma-{1\over2}},1\}
$$
(b) The following values of the constants are allowed: $C_1=24305$, and
$$
C_2(\sigma,r)={C(\sigma)\over\min\{r-\sigma-{1\over2},{7\over8}(2\sigma+1)\}^3},\quad
{\rm with}
$$
$$
C(\sigma)=25223\,\pi(2\sigma+1)^3\left(2(2\sigma+1)\over
e(1-\exp\left(-4\over 2\sigma+1\right))\right)^{\sigma+{1\over2}}.
$$
\vskip1mm
\nid (c) Theorem~1 extends to $V:\R\times\R_+\to\BB(\HH)$ of the type described
above.}
\vsth
In the progressive diagonalization method one must solve at each step a
commutator equation of the type
$$
[K_0+G_s,W_s]=V_s.
$$
This is done block-component wise i.e. with the notation (\ref{componentOfV})
solving for each $(k,m,n)\in\Z\times\N\times\N$ the following matrix equation in
the unknown $W_s(k,m,n)$:
$$
(\omega k+E_m+G_s(m)) W_s(k,m,n)-W_s(k,m,n)(E_n+G_s(n))=V_s(k,m,n).
$$
We are interested in the best possible estimate of
$\|W_s(k,m,n)\|$ in terms of $\|V_s(k,m,n)\|$. In $\S2$ we report on a
method to solve this equation which, we believe, is the best one known so
far. Finally we present two applications in
\S4.

\section{On the commutator equation}

Let $E$ and $F$ be two Hilbert spaces and
$\BB(E)$, $\BB(F)$ the Banach spaces of bounded endomorphisms on $E$ and $F$
respectively, equipped with the usual operator norm. Let $A\in\BB(E)$ and
$B\in\BB(F)$ be selfadjoint and such that
\begin{equation}\label{Condition de separation du spectre}
d_{A,B}:=\dist(\spect(A),\spect(B))>0;
\end{equation}
to each $Y$ in $\BB(F,E)$, the bounded homomorphisms from $F$ into $E$, we want
to associate $X\in\BB(F,E)$ defined as follows:
$$
\ad_{A,B}X=Y,\quad{\rm where}\quad\ad_{A,B}X:=A\,X-X\,B
$$
A review on answers about this question can be found in the beautiful paper
\cite{BhaRos}. In particular one can find there the following result.
\vsth

\nid{\bf Lemma}.

{\sl\ Under the conditions described above
$\ad_{A,B}$ is a bounded linear mapping which has a bounded inverse $\Gamma_{A,B}$
and:
$$
\|\Gamma_{A,B}\|\leq{\pi\over2}\,{1\over d_{A,B}}.
$$
 }

\vsth

\nid{\bf Remark}.

\ (a) In fact there are some special cases when the constant ${\pi\over2}$ can
be replaced by 1. We have not found useful to pay attention to these subtleties
here.
\vskip1mm
\nid (b) The solution $X$ is given by:
$$
X:=\int_\R e^{-itA}Ye^{itB}f(t)dt
$$
with any $f\in L^1(\R)$ such that its Fourier transform $\hat f$ obeys
$\sqrt{2\pi}\hat f(s)=s^{-1}$ on the set $\spect A-\spect B$. Clearly this
shows that $\|X\|\le \|f\|_1\|Y\|$. Optimizing over such $f$ leads to the
constant ${\pi\over2}$.

\section{Unbounded perturbations}

\subsection{The setting}

We start by the description of the class of unbounded perturbations we shall consider.
Let $H_0$ be a {\sl positive selfadjoint} operator on the Hilbert space $\HH$ and
$\{P_n\}_{n\in\N}$ a complete set of mutually orthogonal projections which reduces
$H_0$. We denote by
$E_n:=P_nH_0P_n=H_0P_n$, $\HH_n:=\Ran P_n$ and
$
\HH^{(d)}
$
the algebraic direct sum:
$\oplus_{n\in\N}\Ran P_n$. We introduce the following Banach spaces: for all $1\le
p\le\infty$
$$
L^p(\HH^{(d)})\ni u=\bigoplus_{n\in\N}u_n:\iff
\|u\|_p^p:=\sum_{n\in\N}\|u_n\|^p<\infty
$$
where $\|\cdot\|$ is the norm of $\HH$. Of course $L^2(\HH^{(d)})$ is nothing but
$\HH$ and $\|u\|_\infty:=\sup_n\|u_n\|$.

Then $\BB^{q,p}$, $1\le p,q\le\infty$, will denote the Banach spaces of bounded
operators defined on
$L^p(\HH^{(d)})$ with values in $L^q(\HH^{(d)})$ and $\|\cdot\|_{q,p}$ its
operator norm. We note that
$$
\|X\|_{\infty,1}=\sup_{n,m\in\N}\|X(m,n)\|
$$
and
$$
\|X\|_{1,1}=\sup_{n\in\N}\sum_{m\in\N}\|X(m,n)\|,\quad
\|X\|_{\infty,\infty}=\sup_{m\in\N}\sum_{n\in\N}\|X(m,n)\|
$$
where $X(m,n)$ is the block element of $X$ which acts from $\HH_n$ into $\HH_m$ and
$\|X(m,n)\|$ its norm as a bounded operator in $\HH$. We shall say that
$X\in\BB^{q,p}$ is {\sl symmetric} resp. {\sl antisymmetric} if $X(n,m)=X(m,n)^\star$
resp. $X(n,m)=-X(m,n)^\star$ for all $m,n$. This definition coincides with the usual
one in
$\BB^{2,2}\sim\BB(\HH)$. We remark that if $X$ is symmetric or antisymmetric then
$X\in\BB^{1,1}$ if and only if $X\in\BB^{\infty,\infty}$ if and only if
$X\in\BB_{SH}:=\BB^{1,1}\cap\BB^{\infty,\infty}$; this last operator space is equipped
with the norm $\|X\|_{\rm SH}:=\max\{\|X\|_{1,1},\|X\|_{\infty,\infty}\}$. It is
known, (see \cite[Example III.2.3]{K}) that $\BB_{\rm SH}$ is contained in
all
$\BB^{p,p}$, $1\le p\le\infty$, and in particular in $\BB(\HH)$, and it easy to check
that
$\BB_{\rm SH}$ is a Banach algebra.

On the spectra of $H_0$ we require the two following conditions:
$$
{1\over\Delta E}:=\sup_m\sum_{n\ne m}{1\over
\Delta_{m,n}}<\infty\eqno{(\mbox{GGC$H_0$})}
$$
with
$$ \Delta_{m,n}:=\dist(\spect E_m,\spect E_n)
$$
which expresses that the distances between the spectrum of two blocks $E_m$ and
$E_n$ grows sufficiently rapidly with $|m-n|$. The second condition says that
each blocks $E_n$ must be bounded:
$$
\forall n,\quad E_n\in\BB(\HH).\eqno{(\mbox{BBC$H_0$})}
$$

\subsection{A Class of unbounded perturbations}

We make the following assumptions on the perturbation of $H_0$ to be considered:
$$
V\in\BB^{\infty,1}\ \mbox{\sl and is symmetric}.\eqno{({\rm UV})}
$$
Strictly speaking such a $V$ is not in general an operator acting in $\HH$ but
the following estimate shows that it can be seen as $H_0$-bounded in the
quadratic form sense with zero relative bound: let $R_0(a):=(H_0-a)^{-1}$ with
$a<0$ then
$$
\|R_0(a)^{1\over2}VR_0(a)^{1\over2}\|\le\sum_{n\in\N}{\|V\|_{\infty,1}\over\dist(a,\spect
E_n)}\buildrel {a\to-\infty}\over \longrightarrow 0.
$$
Indeed since that $R_0(a)^{1\over2}$ acts diagonally on $\HH^{(d)}$ one gets
immediately using (GGC$H_0$) that
$$
\max\left\{\|R_0(a)^{1\over2}\|_{1,2},\|R_0(a)^{1\over2}\|_{2,\infty}
\right\}\le\left(\sum_{m\in\N}{1\over\dist(a,\spect
E_m)}\right)^{1\over2}.
$$
This allows to consider $R_0(a)^{1\over2}VR_0(a)^{1\over2}$ as
$$
L^2(\HH^{(d)})\buildrel R_0(a)^{1\over 2}\over \longrightarrow
L^1(\HH^{(d)})\buildrel V\over \longrightarrow
L^\infty(\HH^{(d)})\buildrel R_0(a)^{1\over 2}\over \longrightarrow
L^2(\HH^{(d)});
$$
hence its above estimate and limiting behaviour as $a\to\infty$ follow easily.

\subsection{Progressive diagonalization of $H_0+V$}

Here we show the
\vsth

\nid{\bf Theorem~2}.

{\sl\ Assume $H_0\ge0$ and $V$ obey (GGC$H_0$), (BBC$H_0$) and (UV). If
$$
\|V\|_{\infty,1}\le{\Delta E\over8}
$$
there
exists $J\in\BB_{\rm SH}$ and $G\in\BB^{2,2}$ such that
$$
H_0+V=J(H_0+G)J^\star
$$
with
\nid(i) $[H_0,G]=0$,
\vskip1mm
\nid(ii) $J$ is unitary in $\BB^{2,2}$,
\vskip1mm
\nid(iii) $\|J\|_{\rm SH}\le {3\over2}$ and
$\|G\|\le2\|V\|_{\infty,1}$
\vskip1mm
\nid(iv) $[H_0,J]\in\BB^{\infty,1}$.}
\vsth

\nid{\bf Remark}.

\ (a) Since $\Delta E$ is smaller than the smallest gap of $H_0$ the bound on
$\|G\|\le {\Delta E\over4}$ says in particular that each gap of
$H_0$ remains open after perturbation by $V$. The bound on $J$ will be used
later on.
\vskip1mm
\nid(b) The algorithm says that $G$ belongs to $\BB^{\infty,1}$ which combined
with (i) gives $G\in\BB^{2,2}$.
\vskip1mm
\nid(c) The property (iv) is the key of the so-called "adiabatic regularization
method" first proposed by Howland \cite{H2} for the case of bounded $V$. Its
proof is immediate from the formula $H_0+V=J(H_0+G)J^\star$ since it is
equivalent to $[H_0,J]=JG-VJ$ and since $J\in\BB^{1,1}\cap\BB^{\infty,\infty}$,
$G,V\in\BB^{\infty,1}$. This trick was systematically used in \cite[\S3]{DS}.

\subsubsection{The formal algorithm}

With $H_0+V$ we form a first $4$-tuple of operators
$$\left(U_0:=\id,\quad G_1:=\diag V,\quad
V_1:=\offdiag V,\quad H_1:=H_0+G_1+V_1\right)
$$ where
$$
\diag X:=\sum_{n\in\N}P_nXP_n,\quad\offdiag X:=\sum_{m\ne n}P_mXP_n
$$
Clearly $U_0$ is unitary, $G_1$ diagonal (i.e. commutes with $H_0$), and $V_1$ is
symmetric. Starting from this $4$-tuple we generate recursively an infinite sequence
of such $4$-tuples as follows: let $W_s$ be the solution of
$$
[H_0+G_s,W_s]=V_s\ \&\ \diag\ W_s=0;
$$
we shall use the notations $\ad_AB:=[A,B]:=AB-BA$.
Then we define
\begin{equation}\label{expressionofHsPlus1}
H_{s+1}:=e^{W_s}H_se^{-W_s}=H_0+G_{s}+\sum_{k=1}^\infty{k\over
(k+1)!}\ad_{W_s}^kV_s
\end{equation}
and set
$$
U_s:=e^{W_s} U_{s-1},\quad G_{s+1}:=\diag H_{s+1}-H_0,\quad V_{s+1}=\offdiag
H_{s+1}.
$$
Since $H_0+G_s$ and $V_s$ are symmetric $W_s$ is antisymmetric and therefore
$e^{W_s}$ and $U_s$ are formally unitary. Consequently
\begin{equation}\label{almostDiagonalFormula}
H_0+G_{s+1}+V_{s+1}=U_s (H_0+V) U_s^{-1}
\end{equation}
and to achieve our goal we have to prove that $V_s\to0$, $G_s\to G_\infty$ and
$U_s \to U_\infty$ as $s\to\infty$.

\subsubsection{Convergence of the algorithm}

We solve the commutator equation $[H_0+G_s,W_s]=V_s$ block wise, i.e. for all
$m\ne n$, we look for $W_s(m,n)$ such that
$$
(E_m+G_s(m))W_s(m,n)-W_s(m,n)(E_n+G_s(n))=V_s(m,n).
$$
 Notice the notation
$G_s(m):=G_s(m,m)$. Assume for the moment that
\begin{equation}\label{aprioroconditiononGs}
\forall s\ge1,\ \forall m\in\N,\quad 4\|G_s(m)\|\le\Delta_m
:=\inf_{m\ne n}\Delta_{m,n}
\end{equation}
this implies that $H_0+G_s$ fulfills (BBC$H_0$) and
$$
\forall m\ne n,\quad\dist(\spect E_m+G_s(m),\spect E_n+G_s(n))\ge{1\over2}
\Delta_{m,n}.
$$
Hence by the lemma of \S2 we know that
$W_s(m,n)$ is well defined and obeys
$$
\|W_s(m,n)\|\le
\pi{\|V_s\|_{\infty,1}\over\Delta_{m,n}}\quad\Rightarrow\quad
\|W_s\|_{SH}\le\pi{\|V_s\|_{\infty,1}\over\Delta E}
$$
i.e. $W_s$ belongs to $\BB^{1,1}\cap\BB^{\infty,\infty}$. This shows that
$\ad_{W_s}:\BB^{\infty,1}\to\BB^{\infty,1}$ is bounded by
$2\pi\|V_s\|_{\infty,1}\Delta E^{-1}$ and due to (\ref{expressionofHsPlus1})
$$
\|V_{s+1}\|_{\infty,1}\le\Phi\left(2\pi\|V_s\|_{\infty,1}\over\Delta
E\right)\|V_s\|_{\infty,1}
$$
where $\Phi:\R_+\to\R_+$ is the strictly increasing analytic function defined by
$\Phi(x):=e^x-{1\over x}(e^x-1)$ whose Taylor expansion is
$\sum_{k\ge1}(k/(k+1)!)x^k$.
\begin{center}
\setlength{\unitlength}{0.3mm}
\begin{picture}(100,100)
\put(0,0){\vector(1,0){100}}
\put(98,-8){\small\mbox{$x$}}
\put(0,0){\vector(0,1){100}}
\put(-45,95){\small\mbox{$x\Phi(2x)$}}
\put(0,0){\line(1,1){100}}
\put(47,-8){\mbox{\small$x_\star$}}
\put(-10,50){\mbox{\small${1\over2}$}}
\thicklines
\bezier{0}(0,0)(40, 0)(64,100)
\put(50,0){\line(0,1){1}}
\put(50,2){\line(0,1){0.5}}
\put(50,4){\line(0,1){0.5}}
\put(50,6){\line(0,1){0.5}}
\put(50,8){\line(0,1){0.5}}
\put(50,10){\line(0,1){0.5}}
\put(50,12){\line(0,1){0.5}}
\put(50,14){\line(0,1){0.5}}
\put(50,16){\line(0,1){0.5}}
\put(50,18){\line(0,1){0.5}}
\put(50,20){\line(0,1){0.5}}
\put(50,22){\line(0,1){0.5}}
\put(50,24){\line(0,1){0.5}}
\put(50,26){\line(0,1){0.5}}
\put(50,28){\line(0,1){0.5}}
\put(50,30){\line(0,1){0.5}}
\put(50,32){\line(0,1){0.5}}
\put(50,34){\line(0,1){0.5}}
\put(50,36){\line(0,1){0.5}}
\put(50,38){\line(0,1){0.5}}
\put(50,40){\line(0,1){0.5}}
\put(50,42){\line(0,1){0.5}}
\put(50,44){\line(0,1){0.5}}
\put(50,46){\line(0,1){0.5}}
\put(50,48){\line(0,1){0.5}}
\put(50,50){\line(0,1){0.5}}

\put(0,50){\line(1,0){1}}
\put(2,50){\line(1,0){0.5}}
\put(4,50){\line(1,0){0.5}}
\put(6,50){\line(1,0){0.5}}
\put(8,50){\line(1,0){0.5}}
\put(10,50){\line(1,0){0.5}}
\put(12,50){\line(1,0){0.5}}
\put(14,50){\line(1,0){0.5}}
\put(16,50){\line(1,0){0.5}}
\put(18,50){\line(1,0){0.5}}
\put(20,50){\line(1,0){0.5}}
\put(22,50){\line(1,0){0.5}}
\put(24,50){\line(1,0){0.5}}
\put(26,50){\line(1,0){0.5}}
\put(28,50){\line(1,0){0.5}}
\put(30,50){\line(1,0){0.5}}
\put(32,50){\line(1,0){0.5}}
\put(34,50){\line(1,0){0.5}}
\put(36,50){\line(1,0){0.5}}
\put(38,50){\line(1,0){0.5}}
\put(40,50){\line(1,0){0.5}}
\put(42,50){\line(1,0){0.5}}
\put(44,50){\line(1,0){0.5}}
\put(46,50){\line(1,0){0.5}}
\put(48,50){\line(1,0){0.5}}
\put(50,50){\line(1,0){0.5}}
\end{picture}

\vsth\vsth
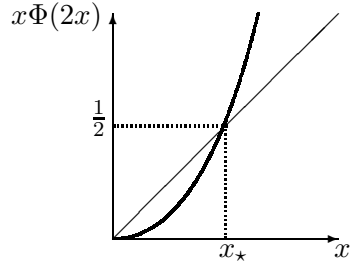
Figure~2. Graph of $x\to x\Phi(2x)$ and its fix point $x_\star={1\over2}$.
\end{center}

\nid With $x_s:=\pi\|V_s\|_{\infty,1}\Delta E^{-1}$, the above inequality
becomes
$x_{s+1}\le\Phi(2x_s)x_s$. This is an elementary exercise to check that the
series $\{x_s\}_s$ is summable if $x_1<x_\star:=1/2$. Thus we get
$$
\|V\|_{\infty,1}\le{\Delta E\over8}\quad\Rightarrow\quad
\sum_{s=1}^\infty
\|W_s\|_{\rm SH}\le\sum_{s=1}^\infty
x_s\le{x_1\over1-\Phi(2x_1)}
$$ The summability of
$\{x_s\}_s$ implies that
$\|V_s\|_{\infty,1}\to 0$ as $s\to\infty$ and that $\sum_{s\ge1}\|W_s\|_{\rm
SH}<\infty$; this last property shows that $U_s$ is convergent in $\BB_{\rm SH}$ to
some $U_\infty$ as $s\to\infty$.

We must check now whether the required property on the
$G_s$, i.e. (\ref{aprioroconditiononGs}), is verified. Since
$G_{s+1}-G_s=\diag\Phi(\ad_{W_s})V_s$ and that $\Delta_m>\Delta E$ we have
successively
\begin{eqnarray*}
(\ref{aprioroconditiononGs})\quad&\Leftarrow&\quad
\sum_{s=1}^\infty\|G_{s+1}-G_s\|+\|G_1\|\le{1\over4}\Delta_m\\
&\Leftarrow&\quad \sum_{s=1}^\infty x_s\Phi(2x_s){1\over\pi}\Delta E+\|G_1\|\le
{1\over4}\Delta E\\
&\Leftarrow&\quad \|G_1\|\le 0.13\,\Delta E\quad\Leftarrow\quad
\|V\|_{\infty,1}\le {\Delta E\over8}
\end{eqnarray*}
since one can check numerically that ${1\over4}-{1\over\pi}\sum_{s=1}^\infty
x_s\Phi(2x_s)\ge 0.13$ if $x_1\le\pi/8$ ( see below for this bound
on
$x_1$). Thus (\ref{aprioroconditiononGs}) is true and we have also shown that
$G_s$ converges to some diagonal and bounded
$G_\infty$ as $s\to\infty$.

To pass from (\ref{almostDiagonalFormula}) to
$H_0+G_\infty=U_\infty(H_0+V)U_\infty^{-1}$ using the three
ingredients $\|V_s\|_{\infty,1}\to0$, $G_s\to G_\infty$ and $U_s\to U_\infty$ is not
as obvious as it seems; we have to adapt the technique of
\cite[\S2.4]{DS}. We have renamed $G_\infty$ by $G$ and $U_\infty$ by $J$ for
later convenience.

Finally we derive the bound on $\|U_\infty\|_{\rm SH}$ and $\|G_\infty\|$.
$$
\| U_\infty\|_{\rm SH}\le\exp(\sum_{k=1}^\infty\|W_k\|_{\rm SH})
\le{3\over2}
$$
since one can check numerically that $\exp(\sum_{s=1}^\infty
x_s\Phi(2x_s))\le {3\over2}$ with
$$
x_1:={\pi\|V_1\|_{\infty,1}\over\Delta E}\le{\pi\over8}.
$$
Concerning $G_s$ notice that $\|X\|_{\rm SH}=\|X\|$ if $X$ is diagonal; then
\begin{eqnarray*}
\|G_\infty\|&\le&\|G_1\|+\sum_{s=1}^\infty\|G_{s+1}-G_s\|
\le
\|V\|_{\infty,1}+\sum_{s=1}^\infty x_s\Phi(2x_s){1\over\pi}\Delta E\\
&\le&\|V\|_{\infty,1}+\left(\Phi(2x_1)+{\Phi(2x_1)\Phi(2x_1\Phi(2x_1))\over
1-\Phi(2x_1\Phi(2x_1)\Phi(2x_1\Phi(2x_1)))}\right)x_1{\Delta
E\over\pi}\\
&=&\|V\|_{\infty,1}\left(1+\Phi(2x_1)+{\Phi(2x_1)\Phi(2x_1\Phi(2x_1))\over
1-\Phi(2x_1\Phi(2x_1)\Phi(2x_1\Phi(2x_1)))}\right)\\
&\le& 2\|V\|_{\infty,1}.
\end{eqnarray*}
The above analytic bound on $\sum_{s=1}^\infty x_s\Phi(2x_s)$ is obtained with
elementary manipulation and we end up with a numerical computation with
$x_1={\pi\over8}$ (notice that $\Phi$ is increasing).

\subsection{Pure pointness of $K_0+V$}

Let $V:\R\to\BB^{\infty,1}$ be a $2\pi$-periodic symmetric function, with the
notation (\ref{componentOfV}) we define the new norm
\begin{equation}\label{normSupOfV}
\tvert V\tvert_r:=\sup_{m,n\in\N}\sum_{k\in\Z}\|V(k,m,n)\|\max\{|k|^r,1\}.
\end{equation}
We shall prove that $K:=K_0+V$ is selfadjoint on a suitable domain and
\vsth

\nid{\bf Theorem~3}.

{\sl\ Let $\omega_0>0$, $\Omega_0:=[{8\over9}\omega_0,{9\over8}\omega_0]$ assume
(\ref{growingGapCondition}) for some $\sigma>0$ and let $\Delta_0:=\min_{m\ne
n}|E_m-E_n|$. Then, $\forall r>\sigma+{3\over2}$, $\exists C_1>0$ and
$\widetilde C_2(\sigma,r)>0$, such that
$$
\tvert V\tvert_{r}<{1\over2(1+8{\omega_0\over\Delta E})}
\min\left\{
{4\Delta_0\over C_1},{\omega_0\over C_1},
{\omega_0\over \widetilde C_2}\left(\Delta E_\sigma\over\omega_0\right)^\sigma,
2^{-r}{\Delta E+8\omega_0\over 4}
 \right\}
$$
implies
$$
\exists\Omega_\infty\subset\Omega_0,\quad {\rm with}\quad
{|\Omega_\infty|\over|\Omega_0|}\ge
1-{2(1+8{\omega_0\over\Delta E})\tvert V\tvert_{r}\over {\omega_0\over
\widetilde C_2}\left(\Delta E_\sigma\over\omega_0\right)^\sigma}
$$
so that $K$ is pure point for all $\omega\in\Omega_\infty$.

In addition one also has that each eigenprojection $P$ of $K$ is bounded in
the norm
$$
\|P\|_{r-\sigma-{3\over2}}=\sup_{m\in\N}\sum_{k\in\Z}\sum_{n\in\N}
\|P(k,m,n)\|\max\{|k|^{r-\sigma-{3\over2}},1\}
$$
and $\widetilde C_2(\sigma,r)=C_2(\sigma,r-1)$, where $C_1$ and $C_2$ are the
constants of Theorem~1. }
\vsth
\nid Proof. (a) As said in the introduction the strategy consists in proving
that $H_0+V(t)=J(t)(H_0+G(t))J^{\star}(t)$ using Theorem~2 for each $t$, then
$$
K_0+V=-i\omega\partial_t+H_0+V=J(-i\omega\partial_t+H_0+\widetilde V)J^\star
$$
where $\widetilde V(t):=G(t)-i\omega J^\star(t)\dot J(t)$ will be seen to
fulfill Theorem~1.

\vskip1mm
\nid(b) Selfadjointness of $K_0+V$ is not an easy matter since the
quadratic form technique cannot be used here because $K_0$ is not bounded
below. We shall establish it indirectly. First with the P.D.M. we shall get the
existence of the strongly $C^1$ map $J: S^1\to\BB_{\rm SH}$ such that $1\otimes
H_0+V=J(1\otimes H_0+G)J^\star$. Then it is easily verified that $K_0+V$ is
selfadjoint on $J\dom K_0$ since $\tilde V$ is bounded.

\vskip1mm
\nid(c) Let $w_r(k):=2^r\max\{|k|^r,1\}$ for some $r\ge0$, we shall use the
notations
$$
w_rV:=\left\{w_r(k)V(k,m,n),k\in\Z,\, m,n\in\N\right\};
$$ it is straightforward to check that
\begin{eqnarray*}
\EE_r&:=&\{V:S^1\to\BB^{\infty,1},\,\tvert w_rV\tvert_0<\infty\} \\
\cA_r&:=&\{V:S^1\to\BB^{\infty,1},\,\| w_rV\|_0<\infty\}
\end{eqnarray*}
are respectively a Banach space and a Banach algebra, with $\cA_r\subset \EE_r$
and
$$
\cA_r\EE_r\subset\EE_r\quad{\rm and}\quad\EE_r\cA_r\subset\EE_r.
$$
We simply follow \S3.3 with $\HH$, $H_0$, $V$, $\BB^{\infty,1}$ and $\BB_{SH}$
replaced respectively by $\KK$, $1\otimes H_0$, $S^1\ni t\to V(t)$, $\EE_r$ and
$\cA_r$ so that we get as for Theorem~2:
\vsth
{\sl if $\tvert w_r V\tvert_0\le\Delta
E/8$ there exists $J\in\cA_r$ and $G\in\EE_r$ such that $1\otimes
H_0+V=J(1\otimes H_0+G)J^\star$ together with
$$
\|w_rJ\|_0\le{3\over2}\quad {\rm and}\quad \|
w_rG\|_0\le2\tvert w_rV\tvert_0.
$$
}
Therefore $\|w_{r-1}J\|_0\le3/4$ and $\|
w_{r-1}G\|_0\le \tvert w_rV\tvert_0$ since $w_{r-1}\le w_r/2$. Of course
it follows that
$\|w_{r-1}J^\star\|_0\le3/4$.

It remains to estimate
$\|w_r\dot J\|_0$. One has with $J=\prod_{s=1}^\infty e^{W_s}$ and
$x_1:=\pi\tvert w_rV\tvert_0\Delta E^{-1}\le\pi/8$:
\begin{eqnarray*}
\|w_{r-1}\dot J\|_0
&\le&\sum_{s=1}^\infty\|w_{r-1}\dot W_s\|_0\exp\left(\sum_{s=1}^\infty\|w_{r-1}
W_s\|_0\right)\\
&=&{1\over2}\sum_{s=1}^\infty\|w_{r}
W_s\|_0\exp\left({1\over2}\sum_{s=1}^\infty\|w_{r} W_s\|_0\right)\\
&\le&{1\over2}
{x_1\over1-\Phi(2x_1)}\exp\left({1\over2}{x_1\over1-\Phi(2x_1)}\right)\\
&\le& \pi{\tvert w_rV\tvert_0\over\Delta E}3=3\pi{\tvert
w_rV\tvert_0\over\Delta E}
\end{eqnarray*}
since one can check numerically that
$(2(1-\Phi(2x_1))^{-1}\exp(x_1/(2(1-\Phi(2x_1)))$ is less than $3$ if
$x_1\le\pi/8$.

Thus we have obtained for all $\omega\in\Omega_0$
\begin{eqnarray*}
2^{r-1}\|\widetilde V\|_{r-1}&=&\|w_{r-1}\widetilde
V\|_0\le\|w_{r-1}G\|_0+{9\over8}\omega_0\|w_{r-1}J^\star\|\|w_{r-1}\dot J\|\\
&\le&\left(1+{9\over8}{3\over4}3\pi{\omega_0\over\Delta E}\right)\tvert
w_rV\tvert_0\le
\left(1+8{\omega_0\over\Delta E}\right)2^r\tvert V\tvert_r.
\end{eqnarray*}
Finally we apply Theorem~1 to $K_0+\widetilde V$ with $r$ replaced by $r-1$ and
$\|V\|_r$ by $2\left(1+8{\omega_0\over\Delta E}\right)\tvert
V\tvert_r$. We also have to impose the additional condition $\tvert w_r
V\tvert_0\le\Delta E/8$. \QED

\section{Applications}

\subsection{The $d$ dimensional quantum top}

Here we give an example of Theorem~1 with unbounded multiplicities of the
spectrum of $H_0$. We consider the model (\ref{pulsedRotor}). $H_0$ is the
Laplace-Beltrami operator on the $d$-dimensional sphere $S^{d}$. Then the
$n^{\rm th}$ eigenvalue obeys
$$
E_n=n(n+d-1)\quad{\rm with}\quad M_n={n+d\choose d}-{n+d-2\choose
d}\buildrel n\to\infty\over\sim {2n^{d-1}\over (d-1)!}
$$
so that the growing gap condition (\ref{growingGapCondition}) is fulfilled if
and only if
$$
\sum_{m>n}{(mn)^{d-1}\over(m^2-n^2)^\sigma}<\infty\iff
\sigma>2d-1.
$$
If $f\in C^s(\R)$ and $W\in C^u(S^{d})$ with
$$
s>r+1>\sigma+{1\over2}+1>2d+{1\over2}\quad{\rm and}\quad u\ge 4
$$
Theorem~1 applies ( see \cite{DLSV} for details). This model has already been
studied by Nenciu in
\cite{N} who found a sufficient condition to rule out the absolutely continuous
spectrum. We have gathered in the next picture what is known so far concerning
this model.

\vskip10mm
\setlength{\unitlength}{0.35mm}
\begin{center}
\begin{picture}(200,200)
\put(0,0){\vector(1,0){200}}
\put(200,-10){\mbox{$d$}}
\put(0,0){\vector(0,1){200}}
\put(-10,200){\mbox{$s$}}
\put(-2,-12){\small\mbox{1}}
\put(40,0){\line(0,1){2}}
\put(38,-12){\small\mbox{}}
\put(80,0){\line(0,1){2}}
\put(78,-12){\small\mbox{2}}
\put(120,0){\line(0,1){2}}
\put(118,-12){\small\mbox{}}
\put(160,0){\line(0,1){2}}
\put(158,-12){\small\mbox{3}}

\put(-8,-2){\small\mbox{2}}
\put(0,40){\line(1,0){2}}
\put(-8,38){\small\mbox{3}}
\put(0,80){\line(1,0){2}}
\put(-8,78){\small\mbox{4}}

\thicklines
\put(0,0){\line(2,1){200}}
\put(0,20){\line(1,1){150}}
\thinlines
\put(160,60){\vector(-1,1){13}}
\put(120,50){\small\mbox{$s=d+1$,
 \quad \cite['97]{N}}}
\put(80,20){\small\mbox{spect$K$ ??}}
\put(155,122){\vector(-1,1){26}}
\put(120,110)
{\mbox{$s=2d+{1\over2}$,
 \quad \cite['01]{DLSV}}}
\put(75,90){\small\mbox{
spect$_{\rm ac}K=\emptyset$}}
\put(10,140){\small\shortstack{
$\spect_{\rm ac}K=\emptyset$\\
and\\
$\spect_{\rm c}K=\emptyset$\\
if\\
$\omega\in\Omega_\infty$, $u\ge4$\\
and $W$ small enough}}
\end{picture}
\vsth\vsth\vsth
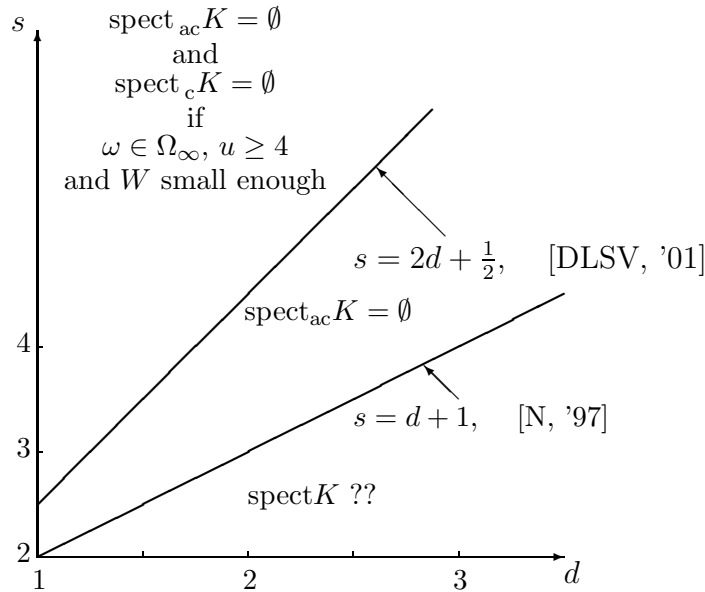
Figure~3. About the quantum top
\end{center}

\subsection{The pulsed rotor with a $\delta$ point interaction}

As an application of Theorem~3 we shall consider the pulsed rotor
(\ref{pulsedRotor}) with
$f\in C^s(S^1)$ and $W$ the delta point interaction located at 0. We
recall that this is the interaction associated to the quadratic form on
$L^2(S^1)$ defined by $u\to |u(0)|^2$. One has for the
$n^{\rm th}$ eigenprojection of
$H_0$
$$
P_n=(\cdot,\varphi_{-n})\varphi_{-n}+(\cdot,\varphi_n)\varphi_n,\quad{\rm
with}\quad
\varphi_n(x):={1\over\sqrt{2\pi}}e^{inx}
$$
except for $P_0=(\cdot,\varphi_0)\varphi_0$. (U$V$) is true since
$\|\delta\|_{\infty,1}=\pi^{-1}$ because
\begin{eqnarray*}
\|P_m\delta P_n\|&=&\left\|{1\over2\pi}\pmatrix{1&1\cr
 1&1\cr}\right\|={1\over\pi}\qquad
m,n\ne0\\
\|P_m\delta P_0\|&=&\left\|{1\over2\pi}\pmatrix{1&1\cr
 }\right\|={1\over\sqrt{2}\pi}\qquad
m\ne0\\
\|P_0\delta P_0\|&=&{1\over2\pi}
\end{eqnarray*}
Moreover
$$
{1\over\Delta E}=\sup_{m\in\N}\sum_{\N\ni n\ne m}{1\over
|m^2-n^2|}={7\over4}\quad\Rightarrow\quad (\mbox{GGC$H_0$})
$$
\vsth
$$
\|E_n\|=\left\|n^2\pmatrix{1&0\cr
 0&1\cr}\right\|=n^2<\infty\quad\Rightarrow\quad
(\mbox{BBC$H_0$}).
$$
Let
$$
f(t)=\sum_{k\in\Z}\hat f_k e^{ikt}
$$
be the Fourier expansion of $f$. Then
$$
\tvert fW\tvert_r\le{1\over\pi}\sum_{k\in\Z}|\hat f_k|\max\{|k|^r,1\}.
$$
Since the eigenvalues of $H_0$ are $\{n^2\}_{n\in\N}$ one has that every
$\sigma>1$ will insure that $\Delta E_\sigma<\infty$. Thus in order to apply
Theorem~3 one needs $r>\sigma+3/2$ i.e. $r>5/2$ and finally $s>7/2$ to
insure that $\tvert fW\tvert_r$ is finite. We have proven that

{\sl\ Let $f\in C^s(\R,R)$ be a $ 2\pi$- periodic function with $s>7/2$ and
$g$ a real constant. The Floquet operator associated to the time dependent
Schr\"odinger operator $-\Delta+gf(\omega t)\delta(x)$ on $L^2(S^1)$ is pure
point provided $g$ is small enough and for appropriate frequencies $\omega$. In
such conditions this quantum system is stable in the sense equation
(\ref{RAGEStability}).}

\vskip 24pt
\noindent{\bf Acknowledgments.}
P.\v{S}. wishes to gratefully acknowledge the partial support from
Grant No. 201/01/01308 of Grant Agency of the Czech Republic. We thank
G.~Burdet and Ph.~Combe for drawing to our attention the review article of
Bhatia and Rosenthal.


\begin{thebibliography}{article}

\bibitem[ADE]{ADE} J.~Asch, P.~Duclos, P.~Exner: {\em Stability of driven
systems with growing gaps, Quantum rings and Wannier ladders}. Journ. Stat.
Phys. {\bf 92}, 1053-, (1998)

\bibitem[BhaRos]{BhaRos}
Bhatia~R., Rosenthal~P.: {\em How to solve the operator equation $AX-XB=Y$}.
Bull. London Math. Soc. {\bf 29} (1997), 1-21

\bibitem[B]{B}
Bellissard~J.: Stability and Instability in Quantum Mechanics in Trend and
development of the eighties, Albeverio and Blanchard. eds, World Scientific,
Singapore, 1985
\bibitem[BaG]{BaG}
Bambusi~D., Graffi~S.:{\em Time quasi periodic unbounded perturbations of
Schr\"odinger operators and KAM methods} Comm. Math. Phys.
{\bf 219} (2001), no. 2, 465--480.

\bibitem[CaG]{CaG}
Casati~G. Guarneri~I.: {\em Non-Recurrent Behaviour in Quantum Dynamics} Comm.
Math. Phys. {\bf 95}, 121-127 (1984)

\bibitem[C]{C}
Combescure M.: {\em The quantum stability problem for time-periodic
perturbations of the harmonic oscillator},
Ann. Inst. Henri Poincar\'e {\bf 47} (1987) 62-82; Erratum:
Ann. Inst. Henri Poincar\'e {\bf 47} (1987) 451-454

\bibitem[DLSV]{DLSV}
P.~Duclos, O.~Lev, P.~\v S\v tov\'\i \v cek, M.~Vittot: {\em Weakly
regular Floquet Hamiltonians with pure point spectrum.}, Rev. Math. Phys. {\bf 14}(6) 2002, pp1-38

\bibitem[DS]{DS}
P.~Duclos, P.~\v S\v tov\'\i\v cek: {\em Floquet Hamiltonians with pure point
spectrum}, Commun. Math. Phys. {\bf 177}, (1996) 327-374

\bibitem[DSV]{DSV}
P. Duclos, P. \v S\v tov\'\i \v cek, M. Vittot: {\em Perturbation of an
eigenvalue from a dense point spectrum: a general Floquet Hamiltonian.}
Ann. Inst. Henri Poincar\'e {\bf71}(3), 241-301, (1999).

\bibitem[EV]{EV}
Enss~V. Veseli\'c~K.: {\em Bound states and propagating states for time-dependent
Hamiltonians}, Ann. Inst. Henri Poincar\'e {\bf 39} (1983) 159--191

\bibitem[H1]{H1}
Howland J. S.: {\em Scattering theory for Hamiltonians periodic in time},
Indiana J. Math. {\bf 28}, (1979) 471-494

\bibitem[H2]{H2}
Howland J. S.: {\em Floquet operators with singular spectrum I, II} Ann. Inst.
Henri Poincar\'e {\bf 49} (1989) 309-334

\bibitem[H3]{H3}
Howland J. S.: {\em Stability of quantum oscillators}, J. Phys. A: Math. Gen.
{\bf 25} (1992) 51777-51181

\bibitem[K]{K}
Kato T.: {\em Perturbation theory of linear operators},
Springer-Verlag, New York, 1966

\bibitem[N]{N}
Nenciu G.: {\em Adiabatic theory: Stability of systems with increasing gaps}
Ann. Inst. Henri Poincar\'e, {\bf 67}(4), (1997) 411--424

\bibitem[Y]{Y}
Yajima K.: {\em Scattering Theory for Schr\"odinger Equations with Potential
Periodic in Time},
J. Math. Soc. Japan {\bf 29} (1977) 729-743
\end{thebibliography}
\end{document}